\documentclass[twocolumn]{aastex701}

\begin{document}

\title{
Constraints on CMBR flux due to high-z dust emission
}

\author[orcid=0000-0001-6128-6274,sname='L\'opez-Corredoira']{Mart\'\i n L\'opez-Corredoira}
\affiliation{Instituto de Astrof\'\i sica de Canarias, E-38205 La Laguna, Tenerife, Spain}
\affiliation{Departamento de Astrof\'\i sica, Universidad de La Laguna, E-38206 La Laguna, Tenerife, Spain}
\email[show]{martin@lopez-corredoira.com}








\begin{abstract}
\citet{Gje25} have proposed a new foreground CMBR component produced by dust associated with the progenitor clouds at $z=15-20$ that led to the formation of massive early-type galaxies, calculating that a minimum of 1.4\%  and a maximum of 100\% of the whole CMBR radiation would be produced by this mechanism.
Here, I check how much dust emission is compatible with the spectrum of the
CMBR monopole flux measurements within the corresponding errors.
{\it COBE}-FIRAS monopole is fitted to different models with different dust emissivity spectral indexes ($\beta $) and dispersion of redshifts of the progenitor clouds ($\sigma _d$).
Within the realistic values of $\beta \ge 0.5$, $\sigma _d\ge 0.8$, the contamination of
CMBR flux by high $z$ dust should be $<1.3$\% (95\% CL).
\end{abstract}

\keywords{\uat{Cosmic microwave background radiation}{322} --- \uat{High-redshift galaxies}{734}
--- \uat{Dust continuum emission}{412}}


\section{Introduction}
\label{.intro}

Dust emission at high $z$ has been proposed several times as a source of the observed cosmic microwave background radiation (CMBR) \citep[e.g.,][]{Ran81,Wic06,Vav18,Mel20},
though it has never been shown to be a realistic solution.

Recently, \citet{Gje25} has revived this idea, claiming that a new foreground CMBR component that has not been previously considered should be produced by the progenitor clouds at $z=15-20$ that led to the formation of massive early-type galaxies, observed with JWST within $z\lesssim 10$ \citep[e.g.,][]{Xia24}.
\citet{Gje25} considered a simple scenario in which the bolometric luminosity of these galaxies at $z\approx 17$ is fully absorbed and thermalized by a high-density dust with $T_{\rm dust}\approx 50$ K permeating the medium without energy loss ($L_{\rm dust}=L_{\rm bol.}$).
At lower redshifts, the dust is rapidly destroyed; therefore, there is little dust
\citep[e.g.,][]{Lab23b},
being the only clue of the existence of this dust their high metallicities.

In this model, the stellar birth of early-type galaxies occurs when the universe was
$\approx 250$ Myr old \citep[Fig. 4/top]{Gje25}, and the peak of maximum bolometric luminosity when the
universe was $\approx 500$ Myr old \citep[Fig. 4/bottom]{Gje25} ($z=9.5$ within $\Lambda $CDM, $z=26$ within $R_h=c\,t$ cosmology; though they place this at $z=17$ assuming a different unspecified cosmological model). In Eq. (27), they attribute at this maximum emission of $z=17$
a total absorption and reemission by dust, but it is not clear whether this dust opacity is maintained at 250 Myr after their stellar birth.

If the numbers calculated by \citet{Gje25} were correct, this foreground would contribute a minimum of 1.4\% of the whole CMBR light, with the possibility of even reaching 100\% depending on the density of early-type galaxies. A 100\% contamination would be very problematic to explain the CMBR power spectrum, since angular size of the observed CMBR fluctuations is much larger than the angular scale of correlation of the high-z galaxies.
Much smaller ratios of contamination might also be incompatible with the CMBR monopole. I will check
it below. 

\section{{\bf COBE}-FIRAS data}

I take the $N=43$ data points of the CMBR monopole $F_{\nu } (\nu)$ within 60 GHz$<\nu <650$ GHz, with the corresponding 1$\sigma $ errors
$\Delta F_{\nu } (\nu)$,
as calculated by \citet{Fix96} from {\it COBE}-FIRAS data. I fit these data with a type
\begin{equation}
\label{model}
F_{\nu; \rm mod.} (\nu)=F_{\nu;\rm CMBR}(\nu)+F_{\nu;\rm dust}(\nu)
,\end{equation}
\[
F_{\nu;\rm CMBR}(\nu)=a_1\,B_\nu(\nu, T_{\rm CMBR})
,\]
\[
F_{\nu;\rm dust}(\nu)=a_2\times
\]\[
\int _0^\infty dz\,P(z)(1+z)\left[\frac{d_L(z_d)}{d_L(z)}\right]^2\,[\nu (1+z)]^\beta \,B_\nu[\nu (1+z), T_{\rm dust}]
,\]
I set a central redshift value of the dust $z_d=17$.
$B_\nu(\nu ,T)$ is a black body radiation, $d_L(z)$ is the luminosity distance.
$\left[\frac{d_L(z_d)}{d_L(z)}\right]\approx \left(\frac{z}{z_d}\right)^{-(1+\alpha )}$
with $\alpha $ depending on the cosmological model
(for instance, $\alpha =0.11$ for $\Lambda $CDM cosmology,
$H_0=70$ km s$^{-1}$ Mpc$^{-1}$, $\Omega _{\Lambda }=0.7$; $\alpha =0.27$ for $R_h=c\,t$ cosmology).

The distribution of redshifts of the clouds' luminosity density is a Gaussian function with dispersion $\sigma _d$.
I try three different values of $\sigma _d=$0,0.8,1.6. In the work by \citet{Gje25},
the cosmic early type galaxy birth function has approximately $\sigma _d=0.8$, so we try here with a maximum dispersion of $2\sigma _d$, and a minimum corresponding to no dispersion ($\sigma _d=0$, i.e. all of the galaxies formed simultaneously at $z=z_d$). I try four different values of dust spectral index
$\beta=0,0.5,1.0,2.0$.
The ratio of FIRAS flux due to dust is:
\begin{equation}
R_{\rm dust}=\frac{\int _{60\ \rm GHz}^{650\ \rm GHz}d\nu \,F_{\nu;\rm dust}(\nu)}
{\int _{60\ \rm GHz}^{650\ \rm GHz}d\nu \,F_{\nu; \rm mod}(\nu )}
.\end{equation}
Table \ref{Tab:dustratio} gives the value of the maximum allowed ratio $R_{\rm dust}$ within 95\% CL for
$\alpha =0.11$ and the twelve different combinations of parameters $\beta $, $\sigma _d$.

\begin{table}
\caption{Best-fitting models of Eq. (\ref{model}), $\alpha =0.11$ with different parameters
$\beta $, $\sigma_d$ to the {\it COBE}-FIRAS monopole \citep{Fix96}.
Range of $R_{\rm dust}$ within 95\% CL, equivalent to $\chi _{red.}^2<2.14\,Min[\chi _{red.}^2]$.
MD stands for the model with the maximum dust contribution [maximum $R_{\rm dust}$ (within 95\% CL)].}
\begin{center}
\begin{tabular}{ccccc}
Dust model  & $T_{\rm CMBR}[{\rm MD}]$ & $T_{\rm dust}[{\rm MD}]$ &
$R_{\rm dust}$ (95\%CL) \\ \hline
No dust & 2.725 & --- & 0 \\ \hline
$\beta =0$, $\sigma _d=0$ & --- & 49 & $<1$ \\
$\beta =0.5$, $\sigma _d=0$ & 2.720 & 53 & $<8.6\times 10^{-3}$ \\
$\beta =1$, $\sigma _d=0$  & 2.715 & 50 &  $<0.012$ \\
$\beta =2$, $\sigma _d=0$  & 2.719 & 44 & $<5.5\times 10^{-3}$ \\\hline
$\beta =0$, $\sigma _d=0.8$ & 2.726 & 49 & $<0.018$ \\
$\beta =0.5$, $\sigma _d=0.8$ & 2.720 & 53 & $<8.3\times 10^{-3}$ \\
$\beta =1$, $\sigma _d=0.8$  & 2.717 & 50 & $<0.010$\\
$\beta =2$, $\sigma _d=0.8$  & 2.719 & 43 & $<6.6\times 10^{-3}$ \\ \hline
$\beta =0$, $\sigma _d=1.6$ & 2.725 & 49 & $<0.011$ \\
$\beta =0.5$, $\sigma _d=1.6$  & 2.725 & 51 & $<5.7\times 10^{-3}$ \\
$\beta =1$, $\sigma _d=1.6$  & 2.720 & 49 & $<6.4\times 10^{-3}$ \\
$\beta =2$, $\sigma _d=1.6$  & 2.718 & 42 & $<6.9\times 10^{-3}$ \\  \hline
\end{tabular}
\end{center}
\label{Tab:dustratio}
\end{table}


The fit for $\beta =0$, $\sigma _d=0$ is compatible with $R_{\rm dust}=1$,
representing the whole radiation coming from a perfect black body with $T_{\rm dust}=T_{\rm CMBR}(1+z_d)$ and without dispersion of redshifts. This however is not realistic because it is utterly impossible that the formation of galaxies takes place exactly at the same time, and a pure black body radiation from dust has never been observed.

More realistic models require $\beta \ge 0.5$, $\sigma _d\ge 0.8$. Within these parameters, the highest value of $R_{\rm dust}$ is 0.010, for $\beta =1$, $\sigma _d=0.8$.
The value of an emissivity index $\beta =1$ is somewhat lower than the usual values for high-$z$ galaxies; \citet{Ben25}
gives for $1.5<z<4.2$ galaxies $\langle \beta \rangle =2.2\pm 0.6$, although some galaxies may reach a value 
$\beta \le 1$.

For these small values of  $\sigma _d$ ($<\frac{z_d}{10}$),
the dependence with $\alpha $ (the dependence with the cosmology)
is negligible, given the symmetry of the Gaussian distribution. If one used $\alpha =0.27$
($R_h =c\,t$) instead of $\alpha =0.11$ ($\Lambda $CDM), one would get the same constraint of maximum
1.0\% (95\% CL) of dust contribution.

Cosmological CMBR anisotropies are negligible in comparison with the residuals of the fits
($\Delta T/T\sim 10^{-4}$).
Dipole was subtracted.
Some questions have been raised about the calibration accuracy of FIRAS, which were addressed by \citet{Fix02}, who also established a more accurate measurement of the monopole of CMBR temperature as $2.725\pm 0.001$ K, in agreement with our best fit without dust of high-$z$ galaxies.

Nonetheless, the monopole obtained by \citet{Fix96} required the subtraction of Galactic (dust) contamination
($G_{\nu ;MW-dust}(\nu )$), and there might be some systematic errors produced by this subtraction.
In order to evaluate the maximum impact in our calculations of this possible systematic errors associated with the Galactic foreground removal, I carry out the following exercise: I assume that the error of each flux point is $\Delta F_{\nu ;\rm FIRAS} (\nu)+|G_{\nu ;MW-dust}(\nu )|$, where the Galactic contribution is the 4th column of \citet[Tab. 4]{Fix96}.
The effect gives slightly larger ratios of $R_{\rm dust}$. For instance, for the case of $\beta =1$, $\sigma _d=0.8$, I get a maximum value of 1.3\% (instead of 1.0\%).
\\



{\bf Acknowledgements:}
Thanks are given to Eda Gjergo and Pavel Kroupa for helpful comments.

\end{document}